# TURNING THE TABLES

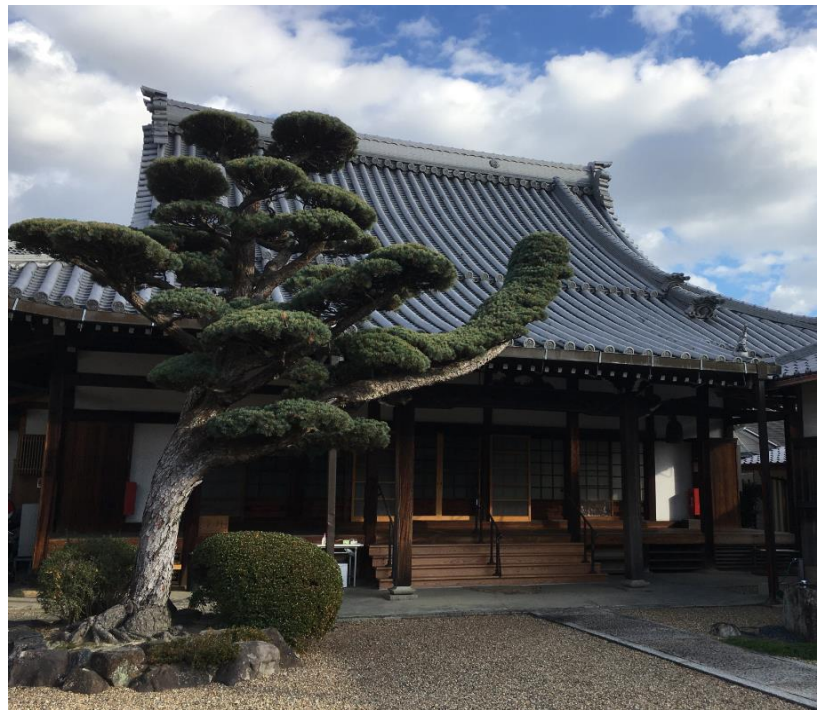

5/10/2021

The View from Offshore During 60 Days in JST

**By James Cusick, PMP**

*j.cusick@computer.org*
*www.researchgate.net/profile/James-Cusick*

---

[1] Cover Photo: Jofukuji Temple (浄福寺), Kawanishi-shi, Hyogo, Japan, 2021

# Turning the Tables

## THE VIEW FROM OFFSHORE DURING 60 DAYS IN JST

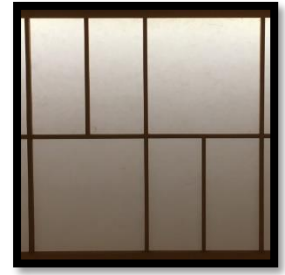

## ABSTRACT

*A report and examination of a Remote Work experience during the Covid-19 pandemic encompassing a 14-hour time difference from the primary work location. Advantages and disadvantages of a globally distributed work experience as compared to an aligned time zone are explored. Logistical aspects of the arrangement are provided as well as the management support, peer reaction, and relative productivity. Recommendations are also provided on how to improve future geographically diverse team arrangements.*

## INTRODUCTION

In late 2020 I began an extended stay in Japan. Due to family obligations I needed to spend time in the region. Fortunately, my employer was willing to consider a temporary flexible work arrangement for me that allowed me to do so for which I have great appreciation. As a result, I essentially became an offshore worker for a 60-day period in the beginning of 2021. During this time, I was working from the JST (Japan Standard Time) time zone but generally aligned my work to the US Eastern time zone to match up with the bulk of my colleague's online availability. Such Global Software Development (GSD) arrangements have been common for many years with my role as manager based in the US and engineering staff located elsewhere [1].

In this paper I will provide some of the details around how this arrangement came about, what it took to pull it off, the advantages of the arrangement as well as the disadvantages. Moreover, I will outline the lessons learned from being an offshore worker even for this short time which I hope to keep in mind as I return to my technical leadership role in the United Sates which includes managing offshore or globally distributed teams from multiple countries. I hope that now I have a better appreciation for their experiences every day and not just through modeling Global Software Development processes as I have done in the past [2].

## BACKGROUND

Once the need for a Remote Work arrangement was established, I discussed the situation with my manager. As my company is focused on information services and my work is in IT (Information Technology), remote work is quite common. Also, during the Covid-19 pandemic, like many companies, physical offices had been generally closed and most staff were working remotely already. So, the primary change for me was around the time zone.

In fact, since I had been a long-term employee and had made many personal trips to Japan in the past, this type of travel was not uncommon in my case and not unknown as related to me by people in my company. Whenever I travelled to Japan on vacation previously, I brought my laptop and was able to keep on top of key correspondence or work items. In one case I recall working on a contract with some team members from the kitchen table in the family home. When I arrived back in New York some of the people involved had not realized that I was on vacation or even out of the country. Thus, precedent did exist for this if not for an extended period.

## ESTABLISHING THE FRAMEWORK

To prepare for an initial discussion with my supervisor I researched the time zone offsets more carefully. I wanted to design a work schedule which I could adhere to in Japan that would maximize my online availability for people in the US. Doing so would reduce the potential for delays in communication or my missing key meetings or people not being able to reach out to me when they needed a question answered.

The first issue with the plan was that New York and Japan were offset by 14 hours. This does vary by an hour when Daylight Savings Time kicks in but that was not going to apply during my stay. As one can see in Figure 1, this is what we call a hard constraint essentially flipping day and night between New York and Japan. The globe is a big place and once you put yourself on the other side of it there are some realities of space and time you cannot defeat. This change in perspective begins to alter the referential center point for planning and has a wide variety of consequences as we shall discuss.

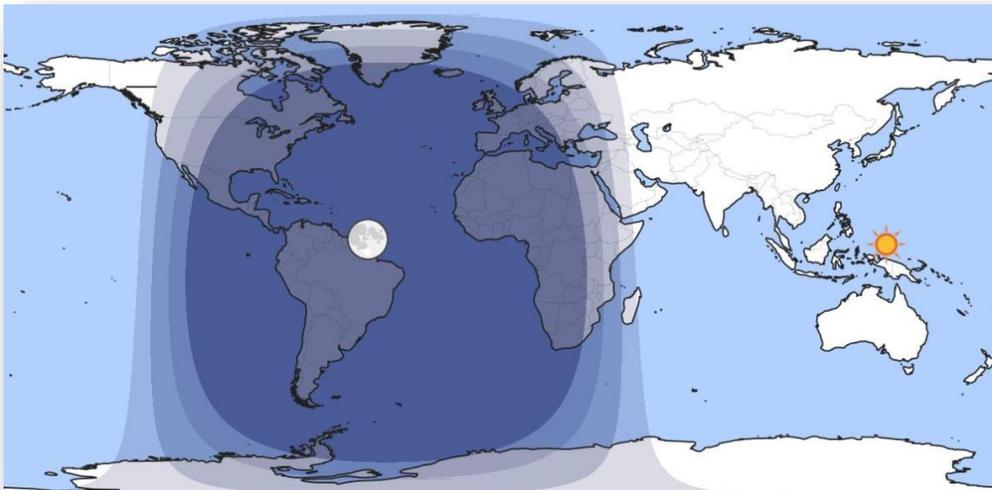

*Figure 1 - Night & day at 11PM EDT March 27, 2021. While NY sleeps it is daytime in Japan [3]*

To further prepare for my discussion with my supervisor I laid out a cross location meeting matrix. Within JST of course there is no variation of having multiple time zones such as in the US, Europe, or other locations. The first step was setting the city where I would be residing (Osaka) and then adding cities with time zones in which there would be primary collaborators. The main area was chiefly New York or the Eastern time zone, Houston, or the Central time zone, followed by India and the Netherlands.

As can be seen in Figure 2, the overlap between JST and EST is minimal. Essentially, the core hours are limited to a few hours in the morning and a few hours in the afternoon of EST. For the Central time zone this gets expanded out a bit. Evening hours in JST line up well with IST and the afternoons for Amsterdam. In reviewing this visual chart and discussing with my supervisor we were able to agree on a plan. My approach would be to work a split shift: a) 6AM JST to 10AM JST, then b) 8PM JST to midnight JST. This would align with a) 4PM EST to 8PM EST, then b) 6AM EST to 10AM EST. In addition, as necessary, I would work offline on project artifacts and communications. In other words, this would be an agreed to framework and I would not be bound only to these hours just as I was not bound to a 40-hour work week as a professional salaried employee located at my home base.

| Osaka | New York | Houston | New Delhi | Amsterdam |
|---|---|---|---|---|
| Tue 6:00 am | Mon 4:00 pm | Mon 3:00 pm | Tue 2:30 am | Mon 10:00 pm |
| Tue 7:00 am | Mon 5:00 pm | Mon 4:00 pm | Tue 3:30 am | Mon 11:00 pm |
| Tue 8:00 am | Mon 6:00 pm | Mon 5:00 pm | Tue 4:30 am | Tue 12:00 midnight |
| Tue 9:00 am | Mon 7:00 pm | Mon 6:00 pm | Tue 5:30 am | Tue 1:00 am |
| Tue 10:00 am | Mon 8:00 pm | Mon 7:00 pm | Tue 6:30 am | Tue 2:00 am |

< Daytime in Osaka >

| Osaka | New York | Houston | New Delhi | Amsterdam |
|---|---|---|---|---|
| Tue 8:00 pm | Tue 6:00 am | Tue 5:00 am | Tue 4:30 pm | Tue 12:00 noon |
| Tue 9:00 pm | Tue 7:00 am | Tue 6:00 am | Tue 5:30 pm | Tue 1:00 pm |
| Tue 10:00 pm | Tue 8:00 am | Tue 7:00 am | Tue 6:30 pm | Tue 2:00 pm |
| Tue 11:00 pm | Tue 9:00 am | Tue 8:00 am | Tue 7:30 pm | Tue 3:00 pm |
| Wed 12:00 midnight | Tue 10:00 am | Tue 9:00 am | Tue 8:30 pm | Tue 4:00 pm |
| Wed 1:00 am | Tue 11:00 am | Tue 10:00 am | Tue 9:30 pm | Tue 5:00 pm |

< Overnight in Osaka >

*Figure 2 - Planned working hours in Osaka as compared to primary colleague's time zones*

## GETTING STARTED

After travelling from NYC to Osaka, which required a number of additional logistical steps due to the ongoing pandemic including meeting various entry and quarantine requirements, I was able to get to work.

### Creating an Office Environment

At first, I set up in the living room where I had worked comfortably in the past. However, I knew this was not going to be sustainable for two months as I simply could not work from the cocktail table. So, we selected a space in the house called the *engawa* (縁側) which is an architectural component of classical Japanese homes. Basically, it is a type of enclosed porch or sunroom with a wood floor and separated off from the main room by *shoji* (the sliding screens most people are familiar with).

After measuring carefully, we did some online shopping and purchased an adjustable standup desk and a comfortable office task chair. We also added a few other items which were in storage in the house like a couple of desk lights, a space heater, and a carpet for the wood floor. I also bought a power strip with a surge protector and ran an extension cable to the work area. Figure 3 provides photographs of both day and night as well as multiple desk height level arrangements.

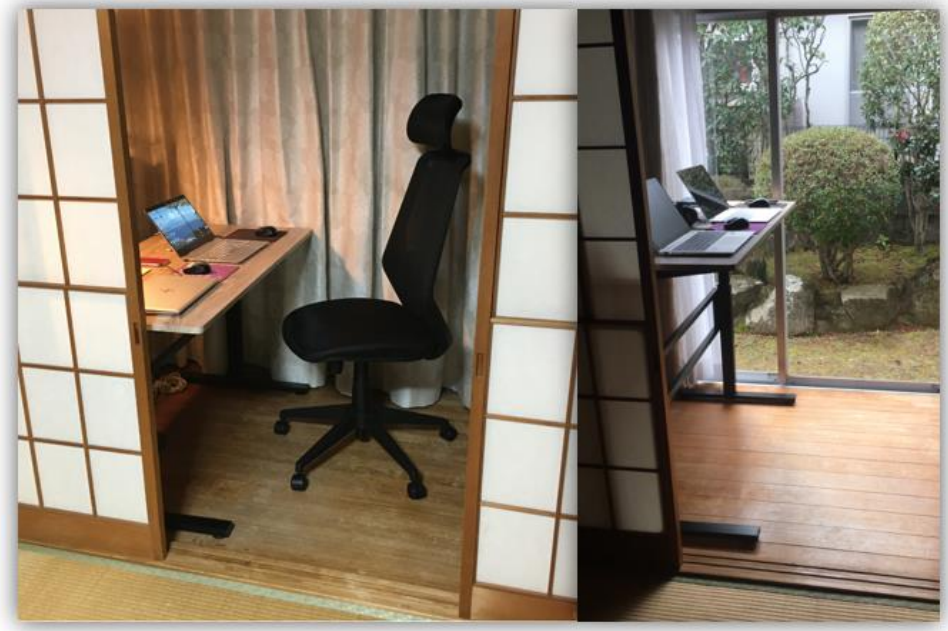

*Figure 3 - The engawa based home office just off formal sitting room separated by shoji doors. Nighttime with desk lowered for sitting (left). Daytime with desk raised for standing (right). Note garden visible from sliding glass doors which made the office area rather pleasant.*

## The Tech

In order to conduct my work, I packed ample gear. I brought with me both my company and personal laptops and cell phones. I had multiple headphones including a pair that were noise cancelling. I also had an external speakerphone, a couple of USB hubs, portable batteries, and memory sticks for backup.

The most critical issue was network access. One thing we did not have in the house was any kind of Internet connectivity as the house is usually unoccupied. As a result, I brought a company provided Mi-Fi device which had worked very well in the past and as a backup my company cellphone which I could use as a hotspot.

As it turns out while these devices could manage email, for example, they did not hold up well to the demands of video conferencing for long. An on-the-fly solution was required. Fortunately, our neighbor who we know well offered their Wi-Fi which I could pick up from my *engawa* office. The bandwidth was somewhat limited, and the signal did drop out occasionally, but this became my primary network solution. I did learn to turn off incoming video on conference calls to conserve bandwidth.

Nevertheless, to ensure connectivity, I ordered a 2GB/Day Portable Wi-Fi (pictured to the right) with a 60-day coverage plan from a company in Tokyo. The unit shipped overnight and was provisioned for immediate use on a domestic carrier provider. With this device I always had a reliable fallback network. While the 2GB limit (provided by 2 1GB SIM cards) was certainly tight, in fact, with the combination of the neighbor's Wi-Fi that was available I never exceeded the maximum use on the portable Wi-Fi when I did use it. My practice was to run the device in the background and switch over to it should the primary Wi-Fi drop out for an extended period.

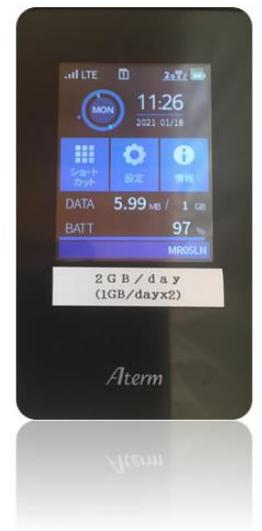

## The Daily Schedule

Once the environment was set up and I began working I kicked into a standard schedule which looked more or less like the below on any given day:

- 5:15AM - wake and eat breakfast.
- 6AM - first morning meeting & begin working.
- 10AM - finish morning work session.
- 10AM to12PM – nap.
- 12PM – lunch.
- 1PM to 6PM – exercise, rest, errands, hobbies, or some work as needed.
- 6PM to 8PM – dinner.
- 8PM – begin evening work session and meetings.
- 12AM or 1AM – conclude evening work session.
- 12AM or 1AM to 5AM – sleep.

*Figure 4 – Portable 2GB/day Wi-Fi Device*

Naturally, this schedule varied from time to time. Primarily, the start or end times might change based on required meetings coming from colleagues or customers. Also, where there were major work products to develop and deliver, I might spend an afternoon working as opposed to resting or running errands.

The most useful tool to manage this scheduling and the complexities of the time zone offsets (which was confusing at first) turned out to be Microsoft Outlook. Within Outlook there is a feature which supports multiple time zones. In this way I was able to configure Outlook with both the JST and EST time zones simultaneously as in Figure 4. This proved invaluable in keeping meeting times straight on a timing basis.

Another key step I took was blocking out my off-duty hours (such as my overnight JST hours) on my calendar. This let other people know that I was not available at those times and they should try to schedule with me at alternative times. This did not always work as some people do not look at such calendar details and schedule regardless of such settings. Nevertheless, this did assist me in planning and in many cases did help others in viewing my availability. Furthermore, I also published a document listing my availability windows to all my teams and collaborators in the company before switching to JST which included the information in Figure 2 above so they could be better prepared for working with me in this new offshore situation.

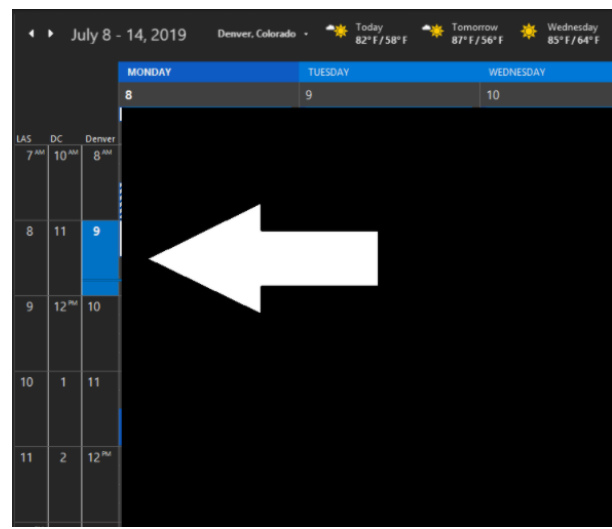

*Figure 5 - Multiple Time Zones in Outlook [4]*

## RESULTS

At the end of the 60 days, we packed our bags and returned to NY. I left my *engawa* office behind for use at some other junction. During those 60+ days we had accomplished the purpose of our stay and at the same time I had maintained my contributions to the company as projected. In fact, I was able to achieve a significant workload and I also learned a lot about being an "offshore" worker. In this section I will detail the benefits and challenges as well as some of the reactions I received from my colleagues.

### Benefits

- The biggest benefit of working from the JST time zone was getting a jump on the day. Ironically, by starting at 8PM JST there was practically no one online back in the US. As a result, I could scan my inbox, draft responses, create documents, post them, and develop new plans. I could do all of this while no one else was reaching out or sending new emails.
- In more than one case, working on the other side of the International Date Line provided an entire day's advantage. For example, my supervisor once requested me to work on an analysis on a Sunday night (EST) which I received on my Monday morning in JST. I spent the day working on it and turned it around completed for his Monday AM EST.
- Another advantage was that since much of my work involves "heads down" research, analysis, and planning, I was able to work independently on such detailed artifacts and deliver them as they were ready for project teams no differently than if I was in the US.
- An advantage of being in JST was its closer alignment with India (IST). There is a good overlap of the two time zones and several times I met with team members in IST when all our colleagues in the US were offline. I was also able to talk live with colleagues in the EU during their afternoon hours. On balance, productivity seemed to be equivalent and sometimes better in this configuration except where direct and protracted collaboration was required with those in EST.
- A surprising benefit of this work schedule was the work/life balance. By working a split shift (a new experience in my career) I was able to utilize a self-programmed afternoon. I used the time for running, taking walks, napping, conducting research, writing, carrying out errands/home repairs, gardening, and even catching up on work. It was a productive time.
- This alternative schedule also allowed for the development of several presentations and papers [5, 6, 7, 8]. I certainly view this output as a major serendipitous benefit of this arrangement. This work covered various topics and I believe the environment provided the right conditions for developing each one.

### Challenges

- The most significant challenge with this arrangement was the schedule itself and the demands it placed on the sleep/wake cycle. Adjusting to operating on about 4-5 hours' sleep took some time especially since I typically sleep much longer than that on average. However, once I established the pattern of the short overnight duration followed by a midday nap, I began to feel much better. In my case I found it easier to work later at night than to wake up earlier in the morning yet that was rarely an option.
- The second most significant issue related to scheduling with colleagues in the US. Since the overlap hours were limited to just a couple of hours in the morning and afternoon and none in the "core hours" of the EST midday people found it difficult to get on my calendar. However, for important sessions we always found a time that could work.
- Unfortunately, no matter how available I made myself and how flexible my team members were, there was a simple fact that events moved fast in NY without me. Sometimes I would come online at 4PM EST after a whole day of meetings had happened in NY and the tactical situation had changed. More than once I had to catch up with events and even sometimes intervene and try to roll back a decision or communicate what had occurred to certain project stakeholders.
- One localized issue we ran into was that on some late-night conference calls my voice would at times carry through the house. I would have to quiet down and remember it was midnight in Japan and not 10AM as it was for others on the call.

## Colleague Reaction

Reactions from colleagues to my offshore work situation ran the gamut. Some people viewed the arrangement with interest and were incredibly positive about it. Others were neutral and just focused on the work. And a few were openly negative as they found the time zone difficult to work with. For a few people they needed some explaining as to how to best engage with me and which times I would be available. Once we did that, most people were able to adjust to the new situation quite well.

In terms of some specific reactions, here is a brief sampling:

- In the best cases, a few people were extremely curious about the culture I was embedded in and wanted to hear details of my environment and asked questions to better understand the experiences I was having. Probably the best line I heard was "How is life in the future", referring to working 14 hours ahead of New York.
- For many people they were simply focused on business matters and never mentioned my remote location whatsoever. They treated me as if I were right next door which I appreciated.
- Some people were completely unaware of the change and did not know that I was even located outside of New York at all. This was true both of some fellow employees and with some customers I support. In the latter case, there was no reason to inform them of my location unless the time zone requirements forced that which it did not often do.
- In one case, when scheduling with a colleague who I worked with frequently, a comment was made that they "did not assume my location". This was somewhat odd as I had advertised my time zone clearly and noted it on every email. Perhaps for some people it remained difficult to take this change into account.
- In another case one colleague abandoned our weekly meetings as the original time needed to be rescheduled during my repositioned period. This was the only case where someone refused to positively adjust to the situation and simply meet at a different time until I returned to the US.
- However, the most direct negative comments from collaborators amounted to - "wish you were in EST" - and "it would be more productive if you were here". Clearly, for some people my remote placement was less than optimal or perhaps their communication style was more open.

In the end, in addition to colleagues, collaborators, and customers I also needed to check in with my supervisor on the effectiveness of this arrangement. At multiple points I asked if I needed to make any adjustments to the approach to make it work better. He said that I was "jumping in at all the right times – so far so good". During the entire period I never had a formal complaint. As a result, I viewed this as positive support from management which goes a long way.

# CONCLUSIONS

This experience yielded several useful lessons. First, one can work effectively from practically any time zone given the right conditions and work responsibilities. Second, doing so may require some sacrifice especially around standard sleep/wake cycles. Third, a radically opposite time zone alignment will reduce on-line collaboration windows for certain time zones but increase collaboration opportunities for other time zones. Once again, the benefits of this are weighted to the work assignment. Fourth, technology plays a key role in enabling such remote work activity. Without the right computing, networking, and communications equipment and tools this arrangement would fail before it begins. Finally, distance truly can disappear in this model. More than once during these 60 days I was on a crystal-clear video call speaking in a full-duplex manner from around the world to people in the US such that they even commented on how it was hard to believe I was speaking with them from Asia.

So, in a time of remote work what does a time zone matter? Effectively, I believe this experience demonstrates that while remote work can be made to function there will be costs and inefficiencies as well. If this experience ran for 600 days instead of 60 days, I am not sure if I would have been able to maintain the daily schedule. Also, the alignment of JST with EST is at such an offset that without some flexibility from the "onshore" personnel (i.e., meeting early AM or late PM[2]), direct communications and collaboration opportunities would remain limited. Finally, with the limited but significant negative reactions witnessed – such as a refusal to meet – I gained true insight into some of the challenges which my colleagues based around the world might be facing daily. Aside from cultural differences and missing the local context this gulf in cooperation and resistance to meet the other party halfway stood out. I recommend any onshore manager to arrange a tour of duty in one of the places they consider offshore.

## REFERENCES

1. Carmel, Erran, **Global Software Teams: Collaborating Across Boarders and Time Zones**, Prentice Hall, Upper Saddle River, NJ, 1999.
2. Cusick, James J., et. al., "A Practical Management and Engineering Approach to Offshore Collaboration", IEEE Software, volume 23, number 5, pages 20--29, 2006, https://doi.org/10.1109/MS.2006.118.
3. Timeanddate.com, https://www.timeanddate.com/worldclock/sunearth.html.
4. "How to Add Multiple Time Zones to Outlook 2019 and Windows 10", January 24, 2020Blog, Microsoft, New Product Releases & Upgrades, Office 365, Office 365, Windows, PEI, https://www.pei.com/add-time-zones-outlook-windows/.
5. Cusick, James, "*Transforming to an AI Rich Work Environment*", **Artificial Intelligence and Its Impacts on The Workforce: A Panel Seminar Session**, The Henry George School of Social Science, New York, January, 2021, [Transforming to AI Rich Work Environments].
6. Cusick, James, & Dragan, Richard, "*Introducing a Land Value Tax with an Online Calculator*", **Eastern Economic Association 47th Annual Conference**, New York, February 25-28, 2021, https://www.researchgate.net/publication/349662201_Introducing_the_Land_Value_Tax_Approach_with_an_Online_Calculator.
7. Cusick, James, "*February Tsubaki (椿) Day by Day: A Photographic Journal of a Camellia bud in late Winter*", March 2021, DOI: 10.13140/RG.2.2.19936.20484/2.
8. Cusick, James, "*What an AI Laundry Machine Taught Me About Economics*", **Tech & Society Blog**, Henry George School of Social Science, 4/27/2021, https://www.hgsss.org/what-an-ai-laundry-machine-taught-me-about-economics/.

---

[2] In one instance along these lines, I half-jokingly suggested to a colleague that since I was working at 6AM JST perhaps they could meet me at 6AM EST once or twice. They burst out in wild laughter at the suggestion.

## ABOUT THE AUTHOR

James Cusick is an interdisciplinary applied researcher specializing in Software Engineering, Cybersecurity, History of Science, and Political Economy. Currently James is Senior Director IT Strategy and Operations & Distinguished Engineer with a global information services firm. Previously James held leadership and technical roles at Dell Professional Services, Bell Laboratories, and AT&T Labs. James currently serves as a Board Trustee at the Henry George School of Social Sciences where he researches topics in Innovation and Economics. James was also Adjunct Assistant Professor at Columbia University's Department of Computer Science. His publications include over 100 articles and papers and two recent books on Information Technology. James is a graduate of both the University of California at Santa Barbara and of Columbia University in New York City. James is a Member of the IEEE Computer Society and a certified PMP (Project Management Professional).